# Controlling and focusing of in-plane hyperbolic phonon polaritons in α-MoO$_3$ with plasmonic antenna

Zebo Zheng[1], Jingyao Jiang[1], Ningsheng Xu[1,2], Ximiao Wang[1], Wuchao Huang[1], Yanlin Ke[1], Huanjun Chen[1*], Shaozhi Deng[1*].

[1]State Key Laboratory of Optoelectronic Materials and Technologies, Guangdong Province Key Laboratory of Display Material and Technology, School of Electronics and Information Technology, Sun Yat-sen University, Guangzhou 510275, China.

[2]Fudan University, Shanghai 200433, China.

*Correspondence to: chenhj8@mail.sysu.edu.cn (H.C.); stsdsz@mail.sysu.edu.cn (S.D.).

**Abstract**

Hyperbolic phonon polaritons (HPhPs) sustained in van der Waals (vdW) materials exhibit extraordinary capabilities of confining long-wave electromagnetic fields to the deep subwavelength scale. In stark contrast to the uniaxial vdW hyperbolic materials such as hexagonal boron nitride (h-BN), the recently emerging biaxial hyperbolic materials such as α-MoO$_3$ and α-V$_2$O$_5$ further bring new degree of freedoms in controlling light at the flatland, due to their distinctive in-plane hyperbolic dispersion. However, the controlling and focusing of such in-plane HPhPs are to date remain elusive. Here, we propose a versatile technique for launching, controlling and focusing of in-plane HPhPs in α-MoO$_3$ with geometrically designed plasmonic antennas. By utilizing high resolution near-field optical imaging technique, we directly excited and mapped the HPhPs wavefronts in real space. We find that subwavelength manipulating and focusing behavior are strongly dependent on the curvature of antenna extremity. This strategy operates effectively in a broadband spectral region. These findings can not only provide fundamental insights into manipulation of light by biaxial hyperbolic crystals at nanoscale, but also open up new opportunities for planar nanophotonic applications.

**Introduction**

Controlling and manipulating electromagnetic waves at the deep subwavelength scale are the key to the constructing compact nanophotonic devices and circuit[1-6]. One of the most intriguing way to achieve this goal is making full use of the polaritons, described as light-matter hybrid quasiparticles arising from the coupling of photons to elementary excitation (such as plasmon, phonon, exciton, etc.) in materials[7-10]. Among these various polaritons, phonon polaritons (PhPs) in polar crystals have attracted much interests because of their ability for extremely confining electromagnetic filed to nanoscale in the infrared to terahertz region[9-12]. In particular, vdW layered materials such as h-BN, $\alpha$-MoO$_3$ have been demonstrated as natural hyperbolic polaritonic materials[13-17]. The hyperbolic phonon polaritons are originated from their extremely anisotropic lattice vibration along different principal axes. In addition to the extremely high confinement and ultra-low optical loss[13,16,18,19], the hyperbolicity of these materials can be incorporated to a variety of nanophotonic applications, such as subwavelength imaging and focusing[20,21], thermal radiation and transfer management[22,23], generating Cherenkov radiation[24,25], defects diagnosis of materials[26], quantum optics[27], and nanoscale energy flow controlling[28-30].

Recently, $\alpha$-MoO$_3$[16,17,31] and $\alpha$-V$_2$O$_5$[32] has been demonstrated as a natural hyperbolic material sustaining HPhPs with extremely anisotropic form, that is, biaxial hyperbolic phonon polaritons, in the mid-infrared to Terahertz region. The polaritons waves propagating in $\alpha$-MoO$_3$ is strongly direction-dependent on its basal plane. Specifically, the isofrequency contour is of hyperbolic geometry[17] (Fig. 1a), which is in stark contrast to the in-plane isotropic HPhPs in other vdW materials such as h-BN, where the in-plane isofrequency curves are circular[20] (Fig. 1b). The in-plane hyperbolicity endows with α-MoO3 an extremely high momentums (towards infinity at specific direction) on its surface, thus bring new avenues to manipulate and configuring infrared light waves and energy flow at the nanoscale flatland. For instance, recent studies have demonstrated the configurable light propagation behaviors in twisted stacking of two α-MoO3 layers[33-36]. The topology of the isofrequency surface can be precisely controlled so that extreme state such as polaritons canalization can be achieved. Furthermore, studies on orientation-dependent highly-confined propagation mode in α-MoO3 cavity[37] and orientation-sensitive near-field radiative heat transfer[23], also have been reported very recently.

To date, subwavelength focusing in hyperbolic h-BN have been extensively studied in previous reports. In specific, previous studies had demonstrated that subwavelength focusing of in-plane isotropic HPhPs in vdW h-BN can be achieved using metallic disk beneath the crystal. However, for HPhPs propagate within α-MoO3 surface with concave wavefront, rather than convex one as those in isotropic materials. The highly in-plane anisotropic dispersion can give rise to orientation-dependent wavevector at a fixed frequency, resulting in major challenge to control and focus these polaritons waves. Conventional concept and structure designs for the in-plane isotropic counterparts, intuitively, may be invalidated for α-MoO3. Theoretically, one can realize in-plane focusing by utilizing the negative refraction across two manually constructed designer metasurfaces, which of vertically orientated optical axis. However, in that configuration, precise fabrication and near-field coupling are needed[38]. Therefore, experimental demonstration based on this strategy has only been realized in microwave region[39]. To the best of our knowledge, the focusing of in-plane hyperbolic polaritons on a

2D basal plane in the mid-infrared region remains elusive so far.

Here, we propose and demonstrate a nontrivial way for the wavefront controlling and focusing of the in-plane HPhPs in α-MoO₃ with geometrically tailored plasmonic antenna. Specifically, by fabricating well aligned metallic nanoantenna with specific extremities, the free space infrared plane waves can be efficiently coupled to the 2D basal plane of α-MoO₃, and subsequently focused to deep subwavelength scale. By combining with numerical simulations, we argue that the in-plane focusing behavior is originated from the synergistical affect between the in-plane hyperbolicity of α-MoO₃ and the appropriate phase arrangement of the antenna.

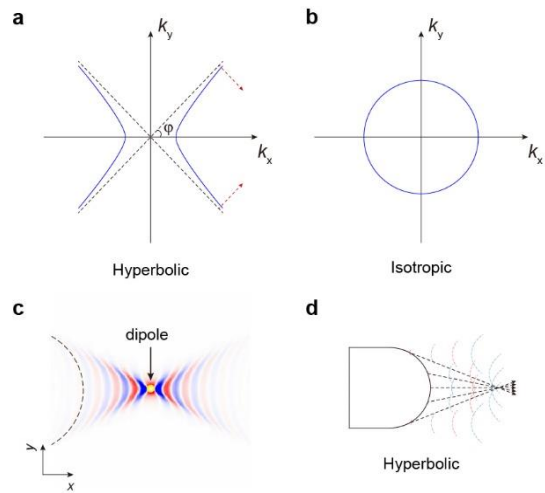

**Fig. 1 In-plane dispersion PhPs in different materials. a** Isofrequency curve of in-plane hyperbolic materials, which is realized in α-MoO₃. **b** Isofrequency curve of in-plane isotropic (out-of-plane hyperbolic) materials, which has been demonstrated in h-BN. **c** Typical concave wavefront launched by a dipole source placed on α-MoO₃ slab with in-plane hyperbolic dispersion. The black dash curve represents the isophase curve of the PhPs. **d** schematic sketch showing the focusing of in-plane hyperbolic waves.

**Results**

**Concept and principle.** In 2D vdW materials sustaining in-plane hyperbolic PhPs, the propagating PhPs waves exhibit hyperbolic isofrequency contour on its basal plane. Specifically, a point dipole source placed at the near-field above such materials can give rise to divergently propagating waves, which is of concave wavefronts (Fig. 1c). The hyperbolic waves can be decomposed as a series of coupled plane waves with varied orientation-dependent wavevectors. To focus the hyperbolic PhPs, one intuitive approach is utilizing concept analogue to the time-reversal technique[40], which is well known in controlling waves in space and time. In a typical time-reversal experiment, a point source is placed at the location where one intends to focus. First, a set of transducers located on a surface are used to record the waves emit from the source. Second, the recorded signals are flipped in time. Third, the flipped signals are launched from the transducers, the resulting wave is found to converge back to its initial source. On the basis of above concept, let us imagine a set of transducers located at an edge with geometry matching the quasi-phase surface of the dipole-launched hyperbolic waves (Fig. 1c). If the transducers can launch waves equivalent to the flipped electromagnetic wave, the superposition of such waves can result in converge wavefront into an infinite focusing spot at the location of the dipole source (Fig. 1d). To achieve such transducers on hyperbolic materials, one should first construct a set of point sources which of phase arrangement matching the concave wavefront of dipole-launched hyperbolic waves. Second, these arranged sources can launch the high-$k$ PhPs by coupling free space plane waves to the near-field. In this scenario, the interference of such waves can give rise to a reproduced point-like spot, thereafter the focusing of HPhPs waves can be achieved (Fig. c,d).

**Controlling and focusing of in-plane hyperbolic PhPs with gold nanoantenna.** For the proof-of-concept demonstration, we fabricated gold antennas using electron beam lithography (EBL) on a large area α-MoO$_3$ single crystal (see Methods). To better demonstrate the controllable PhPs wavefront, the antenna was carefully designed with specific geometry, one of its extremities is of positive curvature (depicted as R1 in Fig. 2b) and another end of negative curvature ((depicted as R2 in Fig. 2b)). The long axis of the antennas was precisely aligned with [100] crystalline direction of α-MoO$_3$ (Fig. 2a and supplementary information, Fig. S1). To verify the control of PhPs wavefront, we image the near-field electromagnetic distribution by using a scattering-type near-field optical microscope (s-SNOM, see Methods)[13,41]. As shown in Fig. 2a, a sharp metal-coated tip (with a curvature of $r$ = 20 nm) embedded in an atomic force microscope (AFM) was illuminated by a focused infrared light (with frequency $\omega_0$). The back-scattered light from the tip was collected using a pseudoheterodyne interferometry and demodulated at higher-order harmonics of the tip vibration frequency. When the sample is scanned beneath the tip, both the topography (Fig. 2b) and near-field optical distribution (Fig. 2c, d) of the sample can be simultaneously obtained.

In our experiment, the focused single frequency laser ($p$-polarized, with focal spot of diameter ~ 30 μm) impact on both the tip and sample with an angle of 40 degree with respect to the sample surface. The laser shine on the sample can be coupled to the nanoantenna, resulting in excitation of localized plasmonic oscillation at the antenna surface[17,34,35,41]. The extremities of the antenna, acting as a secondary source, can excite the propagating PhPs in α-MoO$_3$. The vertical electric component of the PhPs waves can interfere with the tip excited field, forming periodical fringes matching the wavefront of antenna launched PhPs. Fig. 2c shows the near-field optical image obtained at $\omega_0$ = 906 cm$^{-1}$ (Re($\varepsilon_x$) < 0 and Re($\varepsilon_y$), Re($\varepsilon_z$) > 0), which reside in the Reststrahlen band with in-plane hyperbolic dispersion[15-17]. Intriguingly, in the area near the extremity (marked as $R_1$), which with convex semi-circular geometry (curvature $R_1$ = 2), the PhPs waves are launched from the antenna edge and propagates along the [100] direction with a shrinking wavefront. The PhPs waves are focused to a narrow spot and then divergently propagate forward, similar to the free space focusing of plane waves by a conventional convex lens. While in the area near the lower extremity with negative curvature (as depicted in Fig. 2d, $R_2 = -2$), the PhPs waves are in contrary divergent. These nontrivial behaviors are in stark contrast to that in in-plane isotropic materials such as graphene and boron nitride[20,41]. As a direct comparison, we plot the s-SNOM image recorded with $\omega_0 = 998$ cm$^{-1}$ (Fig. 2d), where the PhPs in α-MoO$_3$ are of in-plane elliptical dispersion, similar to that of circular dispersion in in-plane isotropic materials. As shown in Fig. 2d, the PhPs launched by the top upper antenna extremity exhibit a convex wavefront, which approximately follows the geometry of the extremity of antenna. While at the lower extremity, a focusing spot can be observed (Fig. 2d,e), similar to that were previously reported in graphene[41]. We further demonstrate the distinctive observation at the two representative frequencies by plotting the line profiles of the electric filed, $E_z$, of the PhPs wavefronts (Fig. 2e). For the in-plane hyperbolic dispersion ($\omega_0 = 906$ cm$^{-1}$), the lateral size of the focused HPhPs (the can be characterized by the diameter of waist) is ~ 1.3 µm (Fig. 2e, left subpanel), which is ~ 8.5 times smaller than the light wavelength in free space, showing a subwavelength focusing of infared light on the 2D plane. On the contrary, for in-plane isotropic (or approximately isotropic) dispersion ($\omega_0 = 998$ cm$^{-1}$), focusing of PhPs can be enabled by a concave antenna extremity.

The shrinking wavefront at $\omega_0 = 906$ cm$^{-1}$ indicates that the convex antenna extremity, which represents a 2D lens, can focus the in-plane HPhPs whitin the basal plane of α-MoO$_3$ (Fig. 2f). We attribute this non-intuitive behavior to the synergistic effect between the convex plasmonic launcher and the in-plane hyperbolicity of α-MoO$_3$. In contrary to PhPs focusing in the refractive lens, which is governed by the Snell's law[28,42], the gold antenna can firstly couple the far-field infrared plane waves to plasmons at the near-field, the surface confined electromagnetic waves can then be scattered from edge of the antenna and subsequently excite the PhPs in α-MoO$_3$. The experimental observation can be understood by the interference of PhPs waves launched from an array of point sources with specific phase arrangement,

$$\phi(y,\theta) = -k(\theta)[R - \sqrt{R^2 - y^2}]\cos\theta, \quad -R \leq y \leq R, \quad -\pi/2 \leq \theta \leq \pi/2$$

where $y$ represents the position in [001] direction and $R$ is the radius of the antenna extremity. $k$ is the orientation-dependent wavevector of the hyperbolic PhPs waves. According to the Huygens principle, the superposition of PhPs launched by the antenna can be described as,

$$\psi(r_\theta) = \sum_k e^{-i\phi(\theta)} \sum_j C(\theta) \exp(-ik(\theta)r_\theta)$$

Where $C(\theta)$ is the coupling strength of each component with orientation-dependent $k$. The above model shows that the interference of in-plane hyperbolic waves is complicated and intuitive for quantitative demonstration of the beneath mechanism. To simplify the analysis, we corroborate the experimental observations with full-wave electromagnetic simulation (Fig. g,h). In our simulations, the gold antennas set on a α-MoO$_3$ slab were illuminated by perpendicular incident light (see methods). By plotting the $z$-component of electric field distribution on the structure surfaces, the antenna mediated coupling of free space light to the propagating PhPs waves in α-MoO$_3$ can be well reproduced (Fig. g,h).

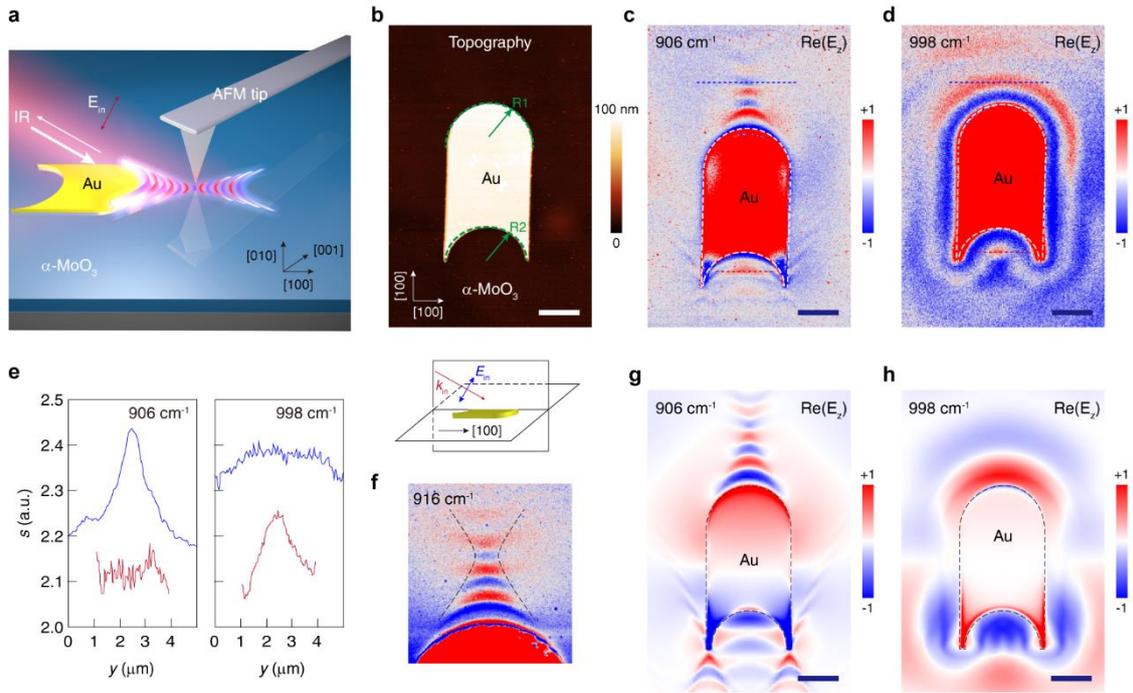

**Fig. 2 Infrared near-field optical nanoimaging of controlled and focused HPhPs. a** Schematic picture of near-field optical measurement using s-SNOM. **b** (Upper) Topographic image of α-MoO$_3$ surface with Au antenna. (Lower) Schematic diagram describes the light illumination in our experiment. **c,d** Near-field optical images of the distinct polaritons waves around the antenna shown in b, with different illumination laser frequencies, 906 cm$^{-1}$ and 998 cm$^{-1}$, respectively. White dash lines marked the edges of the gold antennas. **e** Line plot of near-field optical amplitude along the corresponding blue (red) dash lines in panels **c** and **d**. The gray dash lines represent the Lorentz fitting. **f** Zoom-in s-SNOM image obtained with $\omega_0 = 916$ cm$^{-1}$ for an antenna with $R_1 = 2.5$ μm. **g,h** Simulated Re($E_z$) at frequencies of 906 cm$^{-1}$ and 998 cm$^{-1}$, corresponding to **c** and **d**, respectively. Scale bar in **b-d**, **g**, **h**: 2 μm. Scale bar in **f**: 1 μm.

**Radius of curvature of the antenna extremity.** According to the above discussion, the wavefront of antenna-launched PhPs and the focusing effect at a fixed frequency, are defined by the phase profile, which is determined by the radius of curvature of the antenna extremity. To adequately verify this point, we imaged a set of antennas with varied radius of curvature ($R_1$) at a fixed illumination frequency ($\omega_0$ = 906 m$^{-1}$). As shown in Fig. 3a-d, the antenna-launched PhPs beams keep an identical spatial envelope of hourglass shape while the antennas are of large radius of curvature ($R_1 \geq 1$ μm). The black dash curves represent guides to the eye, indicating the in-plane focusing of PhPs enabled by antennas with varied curvatures. It is noted that, along the principal axis ([100]), the separations of the PhPs fringes, represent the PhPs wavelength ($\lambda_p$), are independent to the extremity curvature (Fig. 3f). The phase profile can only modify the PhPs interference pattern instead of the intrinsic dielectric response of α-MoO$_3$. However, the distances of the PhPs are positively correlated to $R_1$. The focal spot gradually migrates towards the antenna while the radius of curvature decreased. Ultimately, for extremity with small curvature, which is comparable with the PhPs wavelength, the PhPs beam deteriorate to a semi-hourglass envelope with concave wavefront (Fig. 3e), which is consistent to that in antenna launcher with finite lateral size[17,34,35]. We argue that this can be attributed to the ignorable phase difference between secondary sources at the extremity while $R_1$ is small comparable with the PhPs wavelength along [100] direction. The antenna extremity therefore approximately deteriorated to a point source, giving rise to excitation of intrinsic in-plane hyperbolic PhPs.

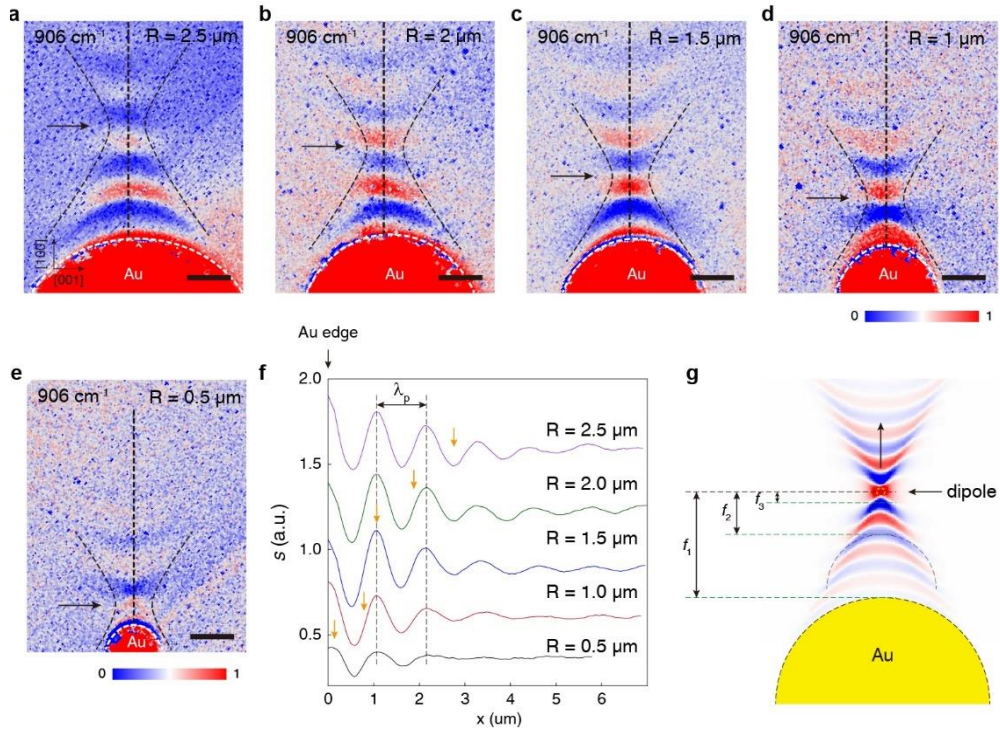

**Fig. 3 Radius of curvature dependent PhPs focusing. a-e** Near-field optical images of the focusing of PhPs with antenna of varied radius of curvatures. The frequency of illumination light is $\omega_0 = 906$ cm$^{-1}$. The black dash curves are guides to the spatial envelop of the PhPs wavefronts. The black arrows indicate the positions of the focal spots. **f** Line profiles along the black dash lines marked at panels (**a-e**). The vertical black dash lines are guide to the eye for the separations of the PhPs osillations. The orange arrows are guide to the eye of the position of focusing spots, corresponding to those are marked by black arrows in panels (**a-e**). **g** Schematic picture of curvature matching between antenna extremities and PhPs wavefronts. Scale bar in (**a-e**) is 1 μm.

We corroborate the above observation by conducting full wave electromagnetic simulations (see Methods section). The simulated near-field electrical distributions are well agreed with the experimentally obtained s-SNOM images in terms of both the fringe separations and curvatures. More importantly, the envelopes of the focusing PhPs change with the radius of curvature of the antenna extremity (supplementary information, Fig. S2). It is noted that the accurate analytical model PhPs may not be intuitive for explanation of the underlying mechanism of the focusing in-plane hyperbolic PhPs. Alternatively, the above observation can be understood by following phenomenological model. Prior to the discussion, we here define the distance between the focal spot and the antenna vertex as the focal length $f$ of the 2D plasmonic lens, analogy to that in conventional lens. Following the discussion on the focusing concept of in-plane hyperbolic PhPs, the antenna that focusing in-plane hyperbolic PhPs can be considered as a time-reversal mirror, which matching the isophase surface of a dipole-launched PhPs. Therefore, a focus spot located at the initial dipole source with infinite lateral size can be reproduced. Fig. 3g shows the transient electric field distribution of a dipole-launched PhPs in α-$MoO_3$. One should note that the propagating PhPs evolves in the space with the wavefronts of changing curvatures, which of increasing radius along with the increased distance from the source. Hence, antenna with different radius of curvature can match the isophase surface at corresponding specific position relative to the source. The excitation of such antennas can launch reverse PhPs therefore recover focusing spots with different focal lengths, which determined by the radius of curvature of the antenna extremities.

It is also notable that the lateral size of the focusing spot should ideally be of infinite small size. However, both the experimental observations and simulation results show that the focus spots are of finite lateral size (Fig. 2e and Fig. 3a-e). We attribute this discrepancy to the slight mismatch of the curvature and the isophase surface of hyperbolic PhPs waves. Furthermore, the relatively higher loss of the high-$k$ components of PhPs can also result in such discrepancy.

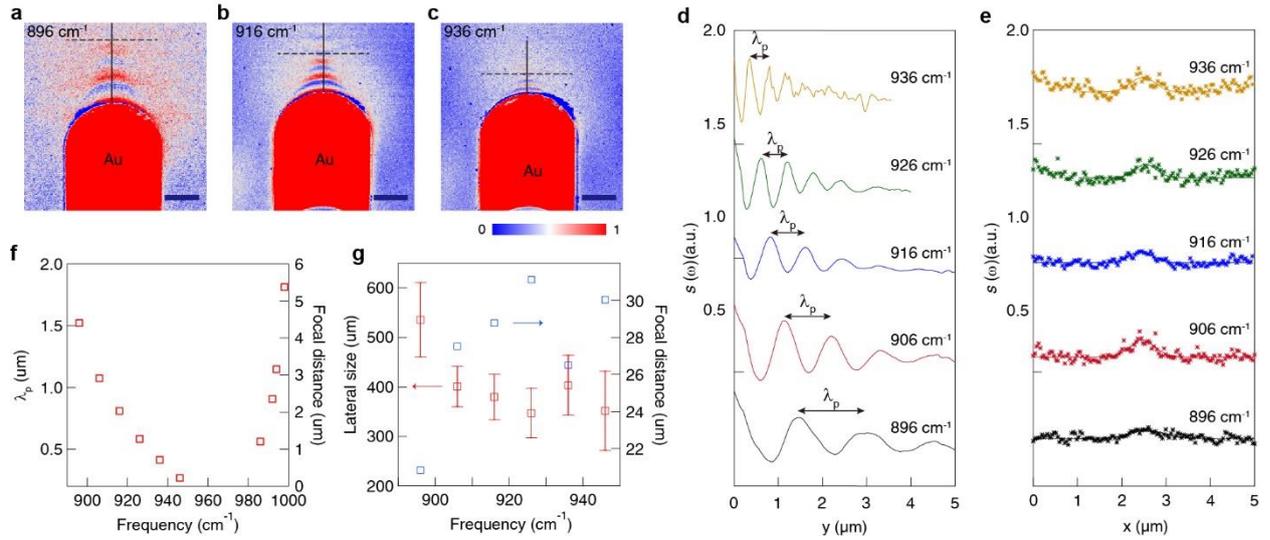

**Fig. 4 Experimental verification of the chromatic aberration of the convex gold antenna. a-c.** Near-field optical images of the antenna ($R_1 = 2.5$ μm) recorded with varied illumination frequencies within the in-plane hyperbolic band. Scale bar: 2 μm. **d.** Line plots of the near-field optical amplitude along the vertical solid lines in (**a-c**). **e** Line plots of the near-field optical amplitude along the horizontal dash lines in (**a-c**). The color dots are experimental data, while the solid lines are Lorentz fitting results. **f** Frequency dependent PhPs wavelengths extracted from the line profiles in (**d**). g Lateral sizes of the focus spots at varied frequencies (red square) and the corresponding focal distances.

**Broadband operation and chromatic aberration.** It is known that the in-plane hyperbolicity of α-MoO$_3$ is preserved within a broad spectral range (545 cm$^{-1}$ − 960 cm$^{-1}$)[17]. However, α-MoO$_3$ is strongly dispersive within this broadband. The PhPs dispersion within this band is mutative, specifically, the opening angle of the isofrequency contour and wavevector magnitude are strongly frequency-dependent. Therefore, it is worthy to evaluate and examine the validity and capability of such antenna designing for focusing of PhPs within the in-plane hyperbolic Reststrahlen band of α-MoO$_3$. To further corroborate the plasmonic antenna mediated in-plane focusing of PhPs in α-MoO₃ in a broad spectral range, we perform s-SNOM imaging with varied excitation frequencies (890 − 1010 cm$^{-1}$) on a representative gold antenna (with extremity of $R_1$ = 2.5 μm) on the α-MoO₃ slab. As shown in Fig. 4a-c, the 2D focused PhPs beam with hourglass shape were observed for representative frequencies within a broadband ranging from 890 cm$^{-1}$ to 946 cm$^{-1}$ (supplementary information, Fig. S3). First, the imaged wavefronts showing periodic fringes with varied spacing along the [100] direction at different frequencies (Fig. 4a-c and Fig. 4d), indicating the effective coupling between incident infrared light and the plasmonic antenna. Furthermore, focus spots of PhPs can be observed at varied frequencies (Fig. 4e). These observations verify that the launching and focusing of in-plane hyperbolic PhPs can be achieved in a broadband infrared spectral range. Second, due to the strong dispersive PhPs response of α-MoO₃, separation and curvature of the PhPs fringes are strongly dependent on light frequency (Fig. 4d), which is well consistent with results reported in many previous studies. Specifically, the PhPs wavelength ($\lambda_p$) varied with the illumination light frequency, showing a frequency dependent electromagnetic confinement (defined as $\lambda_0/\lambda_p$). For example, at $\omega_0$ = 946 cm$^{-1}$, the corresponding $\lambda_p$ is ~267 nm, resulting in an electromagnetic confinement as high as ~ 40.

It is noted that the focal length of a 2D lens is the critical characteristic for practice application such as beam coupling, energy flow management and spectral filtering in planar waveguides and circuits. It is worth discussing the dependence of the focal length on light frequency, namely, the chromatic aberration of the focusing lens. As shown in Fig. 4f, we find that the focal length decreases along with the increasing light frequency (corresponding to decreasing free space wavelength) in the in-plane hyperbolic band (Fig. 4f), giving rise to $f \propto \lambda_p$. Whereas for the in-plane elliptic band, the launched PhPs propagate with radially propagating wavefront (supplementary information, Fig. S4), which is similar to that observed in Fig. 2d. This observation further verifies that the convex antenna extremity for in-plane light focusing operate efficaciously for in-plane hyperbolic PhPs, rather than the in-plane elliptic one. On the other hand, the lateral size of the focal spot is another characteristic for evaluating the operation of the lens. Specifically, in the 2D plane, the size of focal spot represents the lateral confinement of the electromagnetic field. Our s-SNOM imaging results show that PhPs focal spot exhibit deep subwavelength scale within the a broadband region (896 ~ 946 cm$^{-1}$) (Fig. 4e, g). For instance, at $\omega_0$ = 926 cm$^{-1}$, a lateral confinement up to ~36 can be achieved.

**Discussion**

In conclusion, we demonstrated the controlling of in-plane hyperbolic PhPs in α-MoO$_3$ by introducing plasmonic nanoantenna with specific geometry. By combining numerical simulations, we verify that the subwavelength focusing of the hyperbolic PhPs can be achieved with convex antenna extremity, which acts as a 2D lens. In comparison to other counterparts, the Our results can offer novel strategy for subwavelength 2D focusing the infrared light for a variety of planar photonics. The strategy can be applied to essential building blocks for integrated optical systems. For example, nanophotonic devices for on-chip beam splitting, coupling and spectral filtering at the subwavelength scale. The strategy in this study can, in principle, be generalized to other layered materials with in-plane hyperbolic dispersion; thus, a broad range of applications are possible by choosing materials with appropriate hyperbolic bands. Therefore, these results are expected to pave the way toward a new paradigm in manipulating and confining light in planar photonic devices.

**Methods**

**Sample fabrication.** The large area α-MoO$_3$ single crystal (with thickness t = 200 nm) were synthesized by thermal physical deposition method and transferred onto the SiO$_2$/Si substrate. The gold antennas were fabricated onto α-MoO$_3$ using high resolution electron-beam lithography (EBL) followed by evaporation of titanium (5 nm)/gold (100 nm). Afterwards, the photoresist (PMMA) was removed with acetone for 30 minutes.

**Infrared optical nanoimaging.** The optical nanoimaging was conducted using a scattering-type near-field optical microscope (NeaSNOM, Neaspec GmbH). To image the PhPs in real-space, a mid-infrared laser (QCL, Daylight) with tunable wavelengths from 8 μm to 11.2 μm (890 cm$^{-1}$ ~ 1250 cm$^{-1}$) was focused onto both the sample and a metal-coated AFM tip (Arrow-IrPt, Nanoworld). During the measurements, the tip was vibrated vertically with a frequency Ω = 250 kHz. The back-scattered light from the tip was demodulated and detected at a higher harmonic ($\geqslant$ 2Ω) of the tip vibration frequency.

**Numerical simulations.** The full wave simulations were performed using finite element method (COMSOL Multiphysics). The gold antenna on α-MoO$_3$ slab was illuminated by a perpendicularly incident plane wave, with polarization along the long axis of the antenna. The permittivity of Au was taken from Ref. 43. The dielectric functions of SiO$_2$ and α-MoO$_3$ were taken from Ref. 44 and Ref. 45, respectively. The electric field distributions were monitored on the *x-y* plane with vertical distance *z* = 20 nm above the sample surface.

**Acknowledgements**

The authors acknowledge support from the National Natural Science Foundation of China (grant nos. 91963205, 11904420), the National Key Basic Research Program of China (grant nos.

2019YFA0210200, 2019YFA0210203), the Guangdong Natural Science Funds for Distinguished Young Scholars (grant no. 2014A030306017), Guangdong Basic and Applied Basic Research Foundation (grant no. 2019A1515011355, 2020A1515011329). H.C. acknowledge the support from Changjiang Young Scholar Program. Z.Z. acknowledge the project funded by China Postdoctoral Science Foundation (grant no. 2019M663199).

**Author contributions**

H.C., S.D. and Z.Z. proposed the concept and designed the experiments. Z.Z. and J.J. fabricated the structures. Z.Z. performed the optical experiments and numerical simulations. X.W. and Y.K. contributed to the antenna fabrication. W.H. contributed to the growth of α-MoO$_3$ crystal. H.C., S.D., and N.X. coordinated and supervised the work. H.C. and Z.Z. co-wrote the manuscript with the input of all other co-authors.

**Competing interests**

The authors declare no competing interests.

**Additional information**

Supplementary Information accompanies this paper is available at…